# Ultra-high energy cosmic ray investigations by means of EAS muon density measurements

N.S. Barbashina[a], A.G. Bogdanov[a], D.V. Chernov[a], A.N. Dmitrieva[a], D.M. Gromushkin[a], V.V. Kindin[a], R.P. Kokoulin[a], K.G. Kompaniets[a], G. Mannocchi[b], A.A. Petrukhin[a], O. Saavedra[c], V.V. Shutenko[a], D.A. Timashkov[a], G. Trinchero[b], I.I. Yashin[a]

[a] Moscow Engineering Physics Institute (State University), 115409 Moscow, Russia

[b] Istituto Nazionale di Astrofisica, Sezione di Torino, 10133 Torino, Italy

[c] Dipartimento di Fisica Generale dell Universita di Torino, 10125 Torino, Italy

A new approach to investigations of ultra-high energy cosmic rays based on the ground-level measurements of the spectra of local density of EAS muons at various zenith angles is considered. Basic features of the local muon density phenomenology are illustrated using a simple semi-analytical model. It is shown that muon density spectra are sensitive to the spectrum slope, primary composition, and to the features of hadronic interaction. New experimental data on muon bundles at zenith angles from 30° to horizon obtained with the coordinate detector DECOR are compared with CORSIKA-based simulations. It is found that measurements of muon density spectra in inclined EAS give possibility to study characteristics of primary cosmic ray flux in a very wide energy range from $10^{15}$ to $10^{19}$ eV.

## 1. INTRODUCTION

In order to check various theoretical hypotheses concerning the origin, acceleration and propagation of cosmic rays, it is necessary to measure the energy spectrum and composition of cosmic ray particles arriving to the Earth. At energies above $10^{15}$ eV, the only source of information about the primary flux characteristics are extensive air shower (EAS) observations. However, for quantitative interpretation of observation results, an adequate knowledge of the expected EAS characteristics is required. Whereas electromagnetic interactions in EAS are well understood, considerable uncertainties exist in the description of hadronic interactions. The reason is that the yield of secondary particles in the forward kinematic region, which is the most important for phenomenology and understanding of cosmic ray interactions, has not been measured even at energies reached at colliders. Uncertainties of extrapolations to UHE region rapidly increase with the increase of energy. As a consequence, conclusions on primary spectrum and composition inferred from EAS observations appear model dependent. Therefore, careful validation of calculation results on a basis of the comparison with the available data on different EAS observables in a maximal possible range of their variation is important.

In the present paper, experimental data on muon bundles registered at the Earth's surface in a wide range of zenith angles by means of the large area coordinate detector DECOR [1,2] are analysed in terms of a new EAS observable – local muon density at the observation point. Spectra of local muon densities are sensitive to the primary spectrum and composition, and, as it is shown below, are formed by the central part of the showers, thus increasing the sensitivity to the behaviour of most energetic hadrons propagating near the shower axis.

Preliminary data concerning the analysis of muon bundles detected in DECOR in frame of the local muon density approach have been presented elsewhere [3,4]. In particular, it was shown that due to a fast variation of the thickness of the atmosphere and distances from generation level to the observation point with zenith angle the data obtained in this experiment correspond to a very wide interval of primary particle energies. Important advances in the results presented here are the following: increased experimental statistics and new categories of muon bundle events (in multiplicity and zenith angle intervals) have been analysed; a careful consideration of event selection conditions in the detector response calculations has been performed that allowed to substantially decrease systematic uncertainties; the Earth magnetic field (EMF) effects have been included in calculations of the expected distributions.

General scheme used in the present analysis includes the following steps: selection of muon bundle events and construction of distributions in muon bundle multiplicity, zenith and azimuth angles; iterative deconvolution of the measured distributions to detector-independent spectra of local muon density for several zenith angle intervals; simulation of muon lateral distribution functions (LDF) for different types of primary particles, energies, and hadronic interaction models by means of the CORSIKA code [5]; calculation of the expected local muon density distributions by means of the convolution of LDF with a certain primary spectrum and composition model; comparison between experimental and calculation results.

## 2. LOCAL MUON DENSITY PHENOMENOLOGY

As a basis for the analysis of the data on muon bundles, a new phenomenological parameter (local density of muons at the observation point) is used. At large zenith angles, the typical distances of substantial changes of muon lateral distribution function are hundreds – thousands meters, therefore the detector with dimensions of the order of tens meters may be considered as a point-like probe. In a first approximation, the local muon density $D$ (number of muons per unit area) in the event is estimated as the ratio of the number $m$ of muons that hit the detector (muon bundle multiplicity) to the effective detector area $S$ in a given direction. Contribution to the flux of the events with a fixed local density is given by the showers with different primary energies detected at different random distances from the shower axis; however, due to a fast decrease of the cosmic ray flux with the increase of energy, the effective interval of primary particle energies appears relatively narrow.

The main features of the phenomenology of local muon density spectra may be illustrated by means of a following analytical consideration. Without taking into account the fluctuations of the shower development in the atmosphere, the integral spectrum of the events in local density may be written as

$$F(\geq D) = \int N(\geq E(\mathbf{r}, D)) dS ; \qquad (1)$$

here $\mathbf{r}$ is the point in the transverse section of the shower, $N(\geq E)$ is the integral primary energy spectrum, and the minimal energy $E$ is defined by solution of the equation

$$\rho(E, \mathbf{r}) = D , \qquad (2)$$

where $\rho$ ($E,\mathbf{r}$) is muon LDF in a plane orthogonal to the shower axis. As it is seen from the dimensions in Eq.(1), the integral local muon density spectrum is expressed as the number of events per unit solid angle and time interval. The differential density spectrum $dF/dD$ may be easily obtained as derivative of Eq.(1):

$$dF/dD = \int (dN/dE) dS / [d\rho(E,\mathbf{r})/dE] , \qquad (3)$$

where the relation between $E$ and $D$ is again determined by Eq.(2).

Assuming a nearly scaling behaviour of muon LDF around some primary energy $E_0$, it may be parameterized as

$$\rho(E, \mathbf{r}) = (E/E_0)^\kappa \rho(E_0, \mathbf{r}) , \qquad (4)$$

where $\kappa \sim 0.9$. Further, for a power type primary energy spectrum

$$N(\geq E) = A(E/E_0)^{-\gamma} , \qquad (5)$$

we obtain an approximate expression for the local muon density spectrum:

$$F(\geq D) = AD^{-\beta} \int [\rho(E_0, \mathbf{r})]^\beta dS; \beta = \gamma/\kappa . \qquad (6)$$

Thus, in frame of these approximations the spectrum of local muon densities exhibits a power type behavior, a slope being somewhat steeper than that of primary particles. Similar to the spectrum of EAS in the total number of muons $N_\mu$, it increases in absolute intensity for heavier nuclei. The distinctive and important feature of the local muon density spectrum is that it is sensitive to the shape of the muon lateral distribution, since in the integrand of Eq.(6) the LDF enters with index $\beta \sim 2$. Contribution of various distances from the shower axis to the total number of muons and to the local muon density spectrum is illustrated by Fig.1. The abscissa is plotted in the logarithmic scale, therefore respective functions are additionally multiplied by *r*. As it is seen from the figure, the spectrum of local muon density is determined by substantially smaller distances (approximately ten times) than the total number of muons. Thus, selection of the events by muon density enhances the sensitivity of the measured distributions to the central part of the shower. High-energy muons propagating near the shower axis are probes of the early stages of shower development that are governed by high-energy hadronic interactions; therefore data on muon density spectra provide a complementary tool for tests of the validity of hadronic interaction models in the forward region.

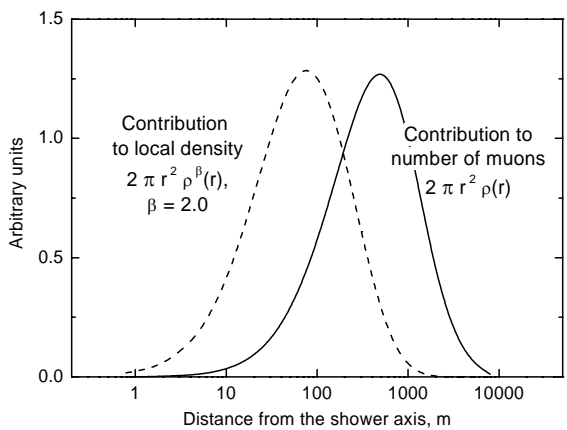

Figure 1. Contribution of various distances to the total number of muons (solid curve) and to the local density spectrum (dashed curve). Muon LDF has been calculated by means of CORSIKA for $10^{17}$ eV primary protons at 60° with hadronic interaction models QGSJET+GHEISHA.

## 3. EXPERIMENTAL

Data collected during long-term experimental runs (14767 hr live time) with the NEVOD-DECOR complex in 2002 – 2005 have been used for the present analysis. A general layout of the setup is shown in Fig.2. The coordinate detector DECOR [1,2] represents a modular multi-layer system of plastic streamer tube chambers with resistive cathode coating, arranged around the Cherenkov water calorimeter NEVOD [6] with sizes $9 \times 9 \times 26$ m$^3$ and a spatial lattice of quasispherical optical modules. The side part of DECOR includes eight 8-layer assemblies (supermodules, SM) of chambers with the total sensitive area 70 m$^2$. Chamber planes are equipped with two-coordinate external strip readout system that allows to localize charged particle track with about 1 cm accuracy in both coordinates (*X*,*Y*). The distance between the adjacent planes of the supermodule is 6 cm. Angular accuracy of reconstruction of muon tracks crossing the SM is better than 0.7° and 0.8° for projected zenith and azimuth angles, respectively. The permanent component of the Earth magnetic field at the setup location is about 52 μT; magnetic inclination is 71° (that is, zenith angle of the magnetic field vector is equal to 19°).

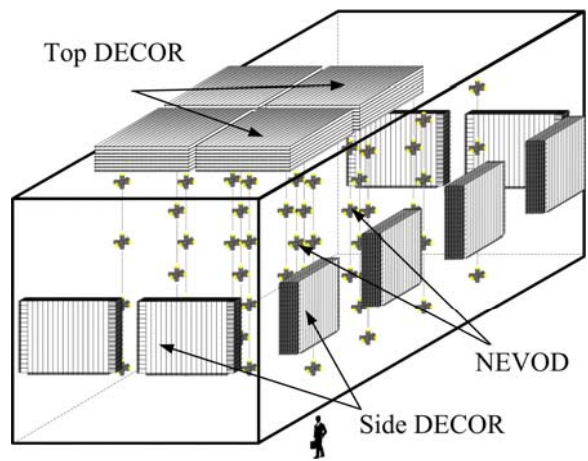

Figure 2. General layout of NEVOD-DECOR complex.

Selection of muon bundle events in the coordinate detector is based on the assumption that tracks of muons produced in the upper atmosphere (far from the setup) are nearly parallel [7]. The selection procedure includes several stages: trigger selection (coincidence of signals from any three supermodules of the side part of DECOR); soft program selection of the events with quasi-parallel (within 5° cone) tracks; scanning of muon bundle candidates, final event classification, and track counting by the operators with the help of a specialised computer interface. Every part of the statistics is independently analysed by at least two operators. After cross-comparison of operators' results, additional examination of controversial events and introduction of necessary corrections, a list of muon bundles containing the bundle multiplicity, zenith and azimuth angle estimates and some auxiliary information is compiled.

At large zenith angles, the EAS reach the setup practically as pure muon component. In these conditions muon bundle events have a very bright signature in the coordinate detector, and their interpretation is unambiguous (Fig.3). However, at lower zenith angles the events are often accompanied by the soft EAS component that complicates geometry reconstruction and muon bundle selection. In order to overcome these difficulties and to extend the range of measurements to moderate zenith angles and low multiplicities (and, correspondingly, to lower primary energies), for zenith angles less than 75° an additional selection cut is applied, namely, only events in two limited sectors of azimuth angle are retained where most of side DECOR supermodules (six of eight) are screened by the water volume of Cherenkov calorimeter, data only of these shielded SM being taken into consideration. Additional filter of 7 – 10 m of water completely absorbs residual electron-photon and hadron components of EAS. Such approach gave possibility to reliably select muon bundles starting from 30° and from a minimal threshold multiplicity ($m \geq 3$) determined by trigger conditions.

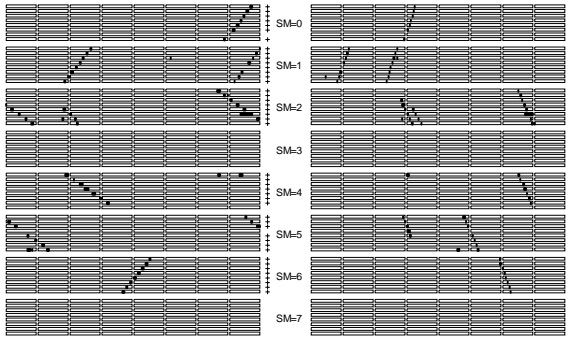

Figure 3. Example of muon bundle event in the coordinate detector (9 parallel tracks, 78° zenith angle). Dark points represent hit strips. Left: $Y$-coordinate strips of SM (azimuth angle measurements); right: $X$-coordinate strips (projected zenith angle).

Since the procedure of muon bundle selection by the operators is rather toilsome, and the number of muon bundles in different multiplicity and zenith angle intervals differs by the orders of magnitude, experimental data are analysed part by part, for separate ranges of $m$ and $\theta$. Statistics of muon bundles of different categories used in the further analysis are summarised in Table 1.

Table 1. Statistics of muon bundle events.

| $m$ | $\theta$ | Time, hr | No. events |
|---|---|---|---|
| $\geq 3$ | 30° – 40° | 250 | 2258 |
| $\geq 3$ | 40° – 60° | 250 | 3748 |
| $\geq 3$ | $\geq 60°$ | 758 | 1928 |
| $\geq 5$ | $\geq 60°$ | 4946 | 3274 |
| $\geq 10$ | $\geq 75°$ | 14767 | 286 |

## 4. DATA ANALYSIS

The procedure of reconstruction of experimental estimates of the local muon density distributions from the observed characteristics of muon bundles is started from the estimation of the parameters of a spectrum model in a following semi-empirical form:

$$dF_0(D,\theta)/dD = C\, D^{-(\beta+1)} \cos^\alpha\theta. \qquad (7)$$

The grounds for a power-type dependence of the local muon density spectrum, at least in some limited range of $D$, are provided by the analytical consideration given in Section 2. On the other hand, preliminary analysis of zenith angle distributions of muon bundles detected in DECOR [8] has shown that these distributions (after introduction of the correction for dependence of the average detector area on zenith angle) are amazingly well described by power function of zenith angle cosine with index $\alpha \sim (4.0 - 4.5)$.

The parameters $\alpha, \beta$ in Eq.(7) are found by means of maximum likelihood method (on the event-by-event basis) for every experimental data sample. The distribution function of the coordinate detector response $P(m,\theta,\varphi;\alpha,\beta)$ for the above model of the spectrum is calculated taking into account the effective setup area $S(\theta,\varphi)$, Poisson fluctuations of the number of muons that hit the detector at a given density $D$, detection efficiency, triggering and selection conditions. Then the expected number of events with a given multiplicity in certain intervals of zenith and azimuth angles $N_{\exp}(m,\Delta\theta,\Delta\varphi)$ is computed. It was found that this model provides quite satisfactory description of the observed distributions in multiplicity and in angles [4].

Finally, experimental estimates of the local muon density spectrum are calculated as the ratios of the observed $N_{obs}$ and expected $N_{exp}$ numbers of the events in a given angular bin, multiplied by the model spectrum Eq.(7) with best-fit parameters, which thus

serves as a first iteration solution for the deconvolution procedure:

$$dF(D,\theta)/dD = [dF_0(D,\theta)/dD] \times$$
$$\times [N_{obs}(m,\Delta\theta,\Delta\varphi)/N_{exp}(m,\Delta\theta,\Delta\varphi)]. \qquad (8)$$

The calculated $dF/dD$ estimates are attributed to certain average values of zenith angle within the corresponding angular intervals and to the average muon density $<D>$ of the events, contributing to the bundles with the observed multiplicity m. Taking into account Poisson fluctuations and the spectrum slope,

$$<D> = (m - \beta) / S_{det}, \qquad (9)$$

where $S_{det}$ corresponds to the average detector area for a given angular range. A reasonable variation of the parameters $\alpha,\beta$ in the spectrum model described by Eq.(7) and used as a first iteration for the deconvolution does not seriously influence the shape and absolute normalisation of the reconstructed spectrum defined by Eq.(8), thus giving the evidence for the robustness of the applied procedure.

## 5. SIMULATION DETAILS

At first, average muon lateral distribution functions have been calculated. For this purpose, simulation of EAS muon component by means of the CORSIKA code [5] (version 6.500) has been performed. Calculations have been done for fixed zenith angles (35°, 50°, 60°, 70°, 80° and 85°), a set of primary energies $E$ (from $10^{14}$ to $10^{19}$ eV, one point per decade), pure protons and pure iron nuclei as primary particles. In the present analysis, combination of hadronic interaction models QGSJET01c + GHEISHA2002 is used; transition from high-energy hadronic interaction model (QGSJET) to the low-energy one (GHEISHA) is made at 80 GeV. The threshold energy of muons and hadrons is set as 2 GeV (close to average muon energy threshold in the experiment); electron-photon component is not simulated. For all variants (except $10^{18}$ and $10^{19}$ eV), the number of simulated events equals to 100; at highest energies, 50 and 20 proton showers, and 20 and 5 showers initiated by iron nuclei have been simulated.

Calculations have been performed in two versions: with consideration of the Earth magnetic field and without it. Main features of the EMF influence on EAS muon component were discussed in [9]. Due to a long distance from muon generation point to observation level, low energy muons are swept out to shower periphery; muons of higher energies are separated in sign and momentum. As a result, the axial symmetry of LDF is destroyed; muon density in the central part of the shower significantly decreases. The last feature is especially impotant in case of detection of muon bundles with a detector of limited area (see discussion in Section 2). The influence of this effect rapidly increases with zenith angle, since the magnetic displacement of particles (in comparison with the case of absense of EMF) is propotional to the squared geometrical path.

In order to obtain two-dimensional muon LDF, particle coordinates given in the CORSIKA output file at a certain horizontal observation level are re-calculated to a plane orthogonal to the shower axis, and average lateral distribution functions $\rho(E,\mathbf{r})$ with a logarithmic step in two coordinates (one of them parallel to the Lorentz force vector, and another one orthogonal to it) are constructed. In further calculations, linear interpolation in ($\log E$, $\log\rho$)-variables is applied.

As a model of the primary flux, a power type all-particle differential spectrum in the form $dN/dE = 5.0 \times (E, \text{GeV})^{-2.7}$ cm$^{-2}$ s$^{-1}$ sr$^{-1}$ GeV$^{-1}$ below the knee energy (4 PeV), steepening to $(\gamma + 1) = 3.1$ above the knee, is used. This spectrum is close to MSU spectrum [10] as given in [11], it is not very much different from the Akeno data [12] around the knee, and is in a reasonable agreement with Fly's Eye "stereo" results [13] around $10^{18}$ eV. To check the sensitivity at highest energies, calculations with the ankle at 3 EeV have been also performed. As two limiting cases of the primary composition, pure proton and pure iron flux are considered. Differential local muon density spectra Eq.(3) at different zenith angles are calculated by means of a convolution with the primary spectrum.

The expected muon density spectra presented in the next section have been calculated with the average LDF (without taking into account fluctuations). Estimatively, consideration of fluctuations in the shower development will increase the expected distributions by 10 – 20 % for primary protons; correction will be less (on a level of few percent) for iron initiated showers.

In Fig.4, results of calculations of the average logarithm of the energy of primary particles that give contribution to the events with a given local muon density $D$ for several zenith angles (labels near the curves) are presented. Calculations have been performed for primary protons, (QGSJET +

GHEISHA) model. The polygons in the figure outline the regions corresponding to selection of muon bundles of different categories (see Table 1). The lower limit of accessible primary energies corresponds to about $10^{15}$ eV because of low muon densities in such EAS. On the other hand, there are statistical limitations around $10^{19}$ eV, since the total number of detected events with muon density $D > 1$ muons/m$^2$ around 80° becomes low.

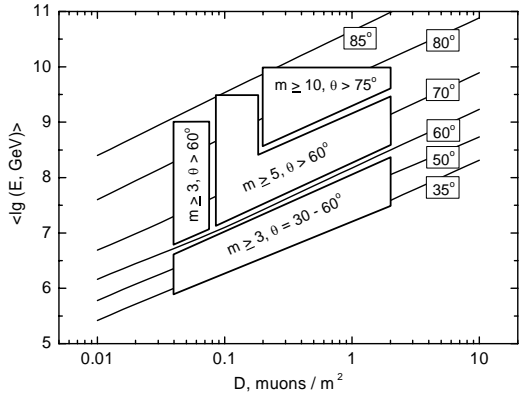

Figure 4. Average logarithms of primary energies, responsible for events with local muon density $D$, for various zenith angles (see the text for details).

## 6. RESULTS AND DISCUSSION

The measured and calculated differential local muon density spectra for zenith angles 35°, 50°, 70°, and 80° are presented in Fig. 5. For the convenience of representation, the spectra are multiplied by $D^3$. The points in the figure are obtained from different sub-sets of the experimental data listed in Table 1; only statistical errors are shown. The curves correspond to calculation results for two extreme versions of primary composition (only protons and only iron) with consideration of the influence of the Earth magnetic field.

A reasonable agreement (including the absolute normalisation) of the present data with CORSIKA-based simulation is observed. At moderate zenith angles, the steepening of the spectra related with the knee is seen. Large multiplicity events in the last angular interval, around 80° (bottom line in Table 1), correspond to energy range $10^{18} – 10^{19}$ eV. Though statistics are limited, a flattening of the muon density spectrum (probably related with the ankle in the primary spectrum) is seen. Data for 70° correspond to intermediate primary energies (about 30 – 100 PeV).

In Fig.6, zenith angular distribution of the events with local muon density exceeding a fixed threshold ($D > 0.04$ muons/m$^2$) is presented as a function of zenith angle cosine (in logarithmic scale). For the comparison, calculation results are presented for two versions: with consideration of the EMF effect and and without it. Difference between these two calculations increases with zenith angle, and at 80 – 85° exceeds the order (!) of magnitude. Experimental data are in a good agreement with simulation performed with EMF. Remarkably, experimental points in Fig.6 lie near a straight line, thus indicating power law dependence of the intensity of the events with a fixed local density threshold on cosθ.

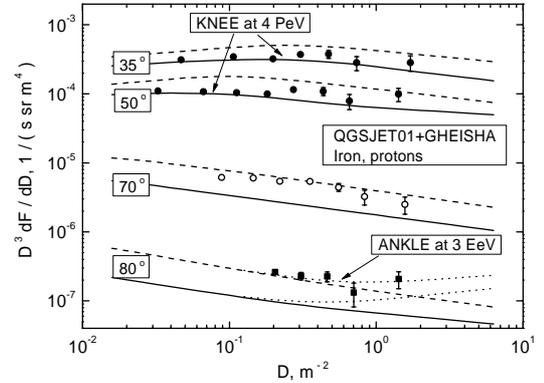

Figure 5. Differential spectra of local muon density for different zenith angles. Points represent different samples of the experimental data; curves are calculation results (dashed and solid curves correspond to primary iron nuclei and protons, respectively). Dotted curves for 80° reflect the expected influence of the ankle.

In Fig.7, angular dependences of the intensity for several threshold densities are compared with simulation results in frame of the model described above; all the curves have been calculated with EMF. Both data and calculations in the figure are divided by a phenomenological factor $\cos^{4.5}\theta$. Within the experimental errors, data for different thresholds exhibit a similar zenith angle dependence. At first sight, comparison of calculated curves and the data in Fig.5 and Fig.7 evidences for the increase of the effective mass of primary particles with the increase of primary energies. It should be stressed however, that all simulations presented here have been performed for a fixed model of the primary spectrum (Section 5), and changes of its slope and normalisation can significantly change the relation between data and the expectation.

Besides, it is necessary to bear in mind also the existing uncertainties in hadronic interaction models at ultra-high energies.

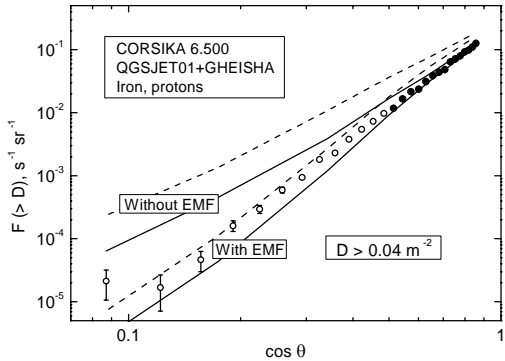

Figure 6. Zenith angle dependence of integral intensity for $D \geq 0.04$ muons/m$^2$. Points correspond to sub-sets of the present data; curves are calculation results with consideration of EMF effect and without it. Notations the same as in Fig.5.

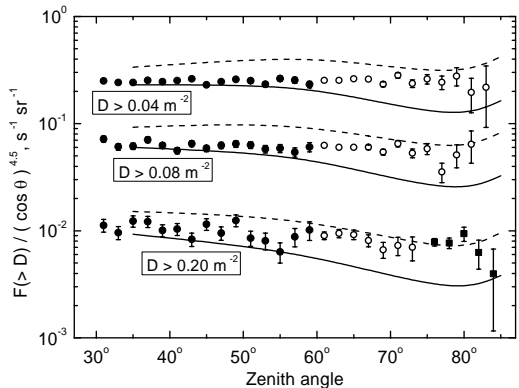

Figure 7. Zenith angular dependence of integral flux of events for several threshold local muon densities. Notations the same as in Fig.5.

## 7. CONCLUSIONS

Analysis of muon bundle events detected at experimental complex NEVOD-DECOR in frame of the approach based on a new phenomenological variable – local muon density – has shown that it is possible to investigate characteristics of primary cosmic ray flux in a very wide energy range (from the knee to the ankle) by means of a single detector of relatively small area. Local muon density distributions are sensitive to the shape of the primary spectrum, primary composition, and have an enhanced sensitivity to a forward region of hadronic interactions.

Comparison of the measured local muon density spectra in a wide range of zenith angles with CORSIKA-based simulations performed with consideration of the Earth magnetic field effects exhibits a reasonable agreement of data with the expectation. For quantitative conclusions, extension of the set of primary flux and hadronic interaction models is required. Hopefully, a further analysis of the data including those of independent experiments will allow to put new constraints on cosmic ray flux and interaction characteristics. The experiment, data analysis and simulations are being continued.

## ACKNOWLEDGEMENTS

The research has been performed at the Experimental Complex NEVOD with the support of Russian Federal Agency for Education, Federal Agency for Science and Innovations, and Russian Foundation for Basic Research (grant no. 06-02-17000a).